\documentclass[12pt]{article}

\def\lsim{\mathrel{\rlap{\lower3pt\hbox{\hskip0pt$\sim$}}
     \raise1pt\hbox{$<$}}}         
\def\gsim{\mathrel{\rlap{\lower4pt\hbox{\hskip1pt$\sim$}}
     \raise1pt\hbox{$>$}}}         

\usepackage{amsmath}
\usepackage{amsfonts}

\begin{document}
\begin{titlepage}

\rightline{LMU-ASC 54/10}

\medskip
\medskip

\centerline{\Large \bf Massive Gravity in de Sitter Space via}
\centerline{\Large \bf Gravitational Higgs Mechanism}
\medskip

\centerline{\large Alberto Iglesias$^\dag$\footnote{\tt Email: alberto.iglesias@physik.uni-muenchen.de} and Zurab Kakushadze$^\S$\footnote{\tt Email: zura@quantigic.com}}

\bigskip

\centerline{\em $^\dag$ Arnold Sommerfeld Center for Theoretical Physics}
\centerline{\em Ludwig Maximilians University}
\centerline{\em Theresienstr. 37, 80333 Munich, Germany}
\medskip
\centerline{\em $^\S$ Quantigic$^\circledR$ Solutions LLC}
\centerline{\em 200 Rector Place, 43C, New York, NY 10280 \footnote{DISCLAIMER: This address is used by the corresponding author for no purpose other than to indicate his professional affiliation as is customary in scientific publications. In particular, the contents of this
paper are limited to Theoretical Physics, have no commercial or other such value, are not intended as an investment, legal, tax or any other such advice, and in no way represent views of Quantigic$^\circledR$ Solutions LLC, the website {\underline{www.quantigic.com}} or any of their other affiliates.}}
\medskip
\centerline{(July 14, 2010)}

\bigskip
\medskip

\begin{abstract}
{}In this paper we discuss massive gravity in de Sitter space via gravitational Higgs mechanism, which provides a nonlinear definition thereof. The Higgs scalars are described by a nonlinear sigma model, which includes higher derivative terms required to obtain the Fierz-Pauli mass term. Using the aforesaid non-perturbative definition, we address appearance of an enhanced local symmetry and a null norm state in the linearized massive gravity in de Sitter space at the special value of the graviton mass to the Hubble parameter ratio. By studying full non-perturbative equations of motion, we argue that there is no enhanced symmetry in the full nonlinear theory. We then argue that in the full nonlinear theory no null norm state is expected to arise at the aforesaid special value. This suggests that no ghost might be present for lower graviton mass values and the full nonlinear theory might be unitary for all values of the graviton mass and the Hubble parameter with no van Dam-Veltman-Zakharov discontinuity. We argue that this is indeed the case by studying full nonlinear Hamiltonian for the relevant conformal and helicity-0 longitudinal modes. In particular, we argue that no negative norm state is present in the full nonlinear theory.

\end{abstract}
\end{titlepage}

\newpage

\section{Introduction and Summary}

{}Gauge interactions are mediated by massless particles, gauge vector bosons. Upon spontaneous breaking of gauge symmetry, gauge bosons acquire mass via Higgs mechanism. Similarly, components of the massless particle associated with general coordinate reparametrization invariance, the graviton, could acquire mass
via gravitational Higgs mechanism \cite{KL} upon spontaneous breaking of diffeomorphism invariance by scalar vacuum expectation values\footnote{For earlier and subsequent related works, see, {\em e.g.}, \cite{Duff, OP, GMZ, Perc, GT, Siegel, Por, AGS, Ch, Ban, AH1, CNPT, AH2, Lec, Kir, Kiritsis, Ber, Tin, Jackiw, RG, HHS}, and references therein.}. The gravitational Higgs mechanism was revisited in the context of obtaining massive gravity directly in four dimensions in \cite{thooft, ZK, Oda, ZK1, ZK2, Oda1, Demir, Cham1, Oda2, JK}. A general Lorentz invariant mass term for the graviton $h_{MN}$ is of the form
\begin{equation}
 -{M^2\over 4} \left[h_{MN}h^{MN} - \kappa (h^M_M)^2\right]~,
\end{equation}
where $\kappa$ is a dimensionless parameter. Unitarity requires that the graviton mass term be of the Fierz-Pauli form with $\kappa = 1$ \cite{FP}. Massive gravity in Minkowski space with $\kappa = 1$ can be obtained via gravitational Higgs mechanism by including higher derivative terms in the scalar sector and appropriately tuning the cosmological constant against the higher derivative couplings \cite{ZK1}.

{}The framework of \cite{ZK1} provides a ghost-free, nonlinear and fully covariant definition of massive gravity in Minkowski space
via gravitational Higgs mechanism with spontaneously (as opposed to explicitly) broken diffeomorphisms. In this paper we discuss massive gravity in de Sitter space via gravitational Higgs mechanism, which provides a nonlinear definition thereof. This is achieved by coupling gravity to scalars, whose vacuum expectation values result in spontaneous breaking of diffeomorphisms, described by a nonlinear sigma model, which includes higher derivative terms required to obtain the Fierz-Pauli mass term.

{}Using our non-perturbative definition of massive gravity in de Sitter space, we discuss the appearance of an enhanced local symmetry and a null norm state in the linearized theory at the special value of the graviton mass $M$ to the Hubble parameter $H$ ratio \cite{DN, Higuchi, DW, GI, GIS}, which in general $D$ dimensions occurs at $M^2 = (D - 2) H^2$. In particular, by studying full non-perturbative equations of motion, we argue that there is no enhanced symmetry in the full nonlinear theory. We then argue that in the full nonlinear theory no null norm state is expected to arise at the aforesaid special value, which in turn suggests that no ghost might be present for $M^2 < (D - 2) H^2$ and the theory might be unitary for all values of $M$ and $H$ with no van Dam-Veltman-Zakharov (vDVZ) discontinuity \cite{vDV,Zak}\footnote{Absence of the vDVZ discontinuity in massive gravity in Minkowski space via gravitational Higgs mechanism was argued in \cite{ZK2}.}. We argue that this is indeed the case by studying full nonlinear Hamiltonian for the relevant conformal and helicity-0 longitudinal modes. In particular, we argue that no negative norm state is present in the full nonlinear theory.

{}The rest of the paper is organized as follows. In Sections 2 and 3 we discuss the gravitational Higgs mechanism in the de Sitter background, which results in massive gravity in de Sitter space with the Fierz-Pauli mass term for the appropriately tuned cosmological constant. In Section 4 we discuss the enhanced local symmetry of the linearized theory and the absence thereof in the full nonlinear theory. In particular, we argue that the full nonlinear theory does not admit solutions obtained by transforming the de Sitter metric via such enhanced local symmetry transformations. In Section 5 we study the full nonlinear Hamiltonian for the relevant conformal and helicity-0 longitudinal modes and argue that no ghost is present for any values of the graviton mass and the Hubble parameter. We briefly summarize our conclusions in Section 6.

\section{De Sitter Solutions}

{}The goal of this section is to obtain massive gravity in de Sitter space via gravitational Higgs mechanism.
Consider the induced metric for the scalar sector:
\begin{equation}
 Y_{MN} = Z_{AB} \nabla_M\phi^A \nabla_N\phi^B~.
\end{equation}
Here $M = 0, \dots, (D-1)$ is a space-time index, and $A = 0,\dots, (D-1)$ is a global index.
We will choose the scalar metric $Z_{AB}$ to be conformally flat de Sitter metric:
\begin{equation}
 Z_{AB} = \omega^2(\phi)~\eta_{AB}~,
\end{equation}
where
\begin{equation}
 \omega(\phi) \equiv {\nu \over n_A \phi^A}~,
\end{equation}
$\nu$ is a dimensionless coupling, and $n_A \equiv (1,0,\dots,0)$. Also, let
\begin{equation}\label{Y}
 Y\equiv Y_{MN}G^{MN}~.
\end{equation}
The following action, albeit not the most general\footnote{In the de Sitter as well as Minkowski cases one can consider a more general setup where the scalar action is constructed not just from $Y$, but from $Y_{MN}$, $G_{MN}$ and $\epsilon_{M_0\dots M_{D-1}}$, see, {\em e.g.}, \cite{ZK1,Demir,Cham1,Oda2}. However, a simple action containing a scalar function $V(Y)$ suffices to capture all qualitative features of gravitational Higgs mechanism. In particular, if this function is quadratic as in (\ref{quad.pot}) the cosmological constant $\Lambda$ must be negative in the context of Minkowski background (but not in the de Sitter case - see below), generically there is no restriction on $\Lambda$, which can be positive, negative or zero even in the context of the Minkowski background, once we allow cubic and/or higher order terms in $V(Y)$, or consider non-polynomial $V(Y)$.}, will serve our purpose here:
\begin{equation}
 S_Y = M_P^{D-2}\int d^Dx \sqrt{-G}\left[ R - V(Y)\right]~,
 \label{actionphiY}
\end{equation}
where {\em a priori} $V(Y)$ is a generic function of $Y$.

{}The equations of motion read:
\begin{eqnarray}
 \label{phiY}
 && \nabla^M\left(V^\prime(Y)Z_{AB} \nabla_M \phi^B\right) = {\partial\ln(\omega)\over \partial\phi^A}~Y V^\prime(Y)~,\\
 \label{einsteinY}
 && R_{MN} - {1\over 2}G_{MN} R = V^\prime(Y) Y_{MN}
 -{1\over 2}G_{MN} V(Y)~,
\end{eqnarray}
where prime denotes derivative w.r.t.~$Y$. Multiplying (\ref{phiY}) by $Z_{AB} \nabla_S\phi^B$ and contracting indices, we
can rewrite the scalar equations of motion as follows:
\begin{equation}\label{phiY.1}
 \partial_M\left[\sqrt{-G} V^\prime(Y) G^{MN}Y_{NS}\right] - {1\over 2}\sqrt{-G} V^\prime(Y) G^{MN}\partial_S Y_{MN} = 0~.
\end{equation}
Since the theory possesses full diffeomorphism symmetry, (\ref{phiY.1}) and (\ref{einsteinY}) are not all independent but linearly related
due to Bianchi identities. Thus, multiplying (\ref{einsteinY}) by $\sqrt{-G}$, differentiating w.r.t.~$\nabla^N$ and contracting indices we arrive
at (\ref{phiY.1}).

{}We are interested in finding solutions of the form:
\begin{eqnarray}\label{solphiY}
 &&\phi^A = m~{\delta^A}_M~x^M~,\\
 \label{solGY}
 &&G_{MN} = \exp(2A)~\eta_{MN}~,
\end{eqnarray}
where $m$ is a mass-scale parameter, and $A = A\left(n_M x^M\right)$ only depends on the time coordinate. The scalar equations of motion (\ref{phiY}), or equivalently (\ref{phiY.1}), then imply that
\begin{equation}
 A = -\ln\left( H n_M x^M\right)~,
\end{equation}
where $ H$ is the constant Hubble parameter. Furthermore, (\ref{einsteinY}) implies that
\begin{equation}
 R_{MN} - {1\over 2}G_{MN} R = -{1\over 2}G_{MN} {\widetilde \Lambda}~,
\end{equation}
where
\begin{equation}\label{cosm.const}
 {\widetilde \Lambda} \equiv V(Y_*) - {2\over D}~Y_* V^\prime(Y_*)
\end{equation}
is the cosmological constant for the de Sitter space whose metric is given by (\ref{solGY}), and we have:
\begin{eqnarray}
 &&Y_* = D\nu^2 H^2 \equiv D\mu^2~,\\
 && H^2 = {{\widetilde \Lambda}\over (D-1)(D-2)}~.
\end{eqnarray}
Thus, the solution is completely determined by the ``potential'' $V(Y)$ and the coupling $\nu$ as $Y_*$ is a solution of the following equation:
\begin{eqnarray}\label{Y_*}
 Y_* = {\nu^2\over (D-1)(D-2)} \left[ D V(Y_*) - 2 Y_* V^\prime(Y_*) \right]~.
\end{eqnarray}
Note that, since ${\widetilde \Lambda}$ must be positive, the potential $V(Y)$ cannot be completely arbitrary. Also, $Y_*$ and ${\widetilde \Lambda}$ are independent of $m$ in (\ref{solphiY}), which is due to the invariance of $Y_{MN}$ under simultaneous global rescalings $\phi^A \rightarrow \lambda\phi^A$.

\section{Massive de Sitter Gravity}

{}In this section we study linearized fluctuations in the background given by (\ref{solphiY}) and (\ref{solGY}). Since diffeomorphisms are broken spontaneously, the equations of motion are invariant under the full diffeomorphism invariance. The scalar fluctuations $\varphi^A$ can therefore be gauged away using the diffeomorphisms:
\begin{equation}\label{diffphiY}
 \delta\varphi^A =\nabla_M \phi^A \xi^M = m~{\delta^A}_M ~\xi^M~.
\end{equation}
However, once we gauge away the scalars, diffeomorphisms
can no longer be used to gauge away any of the graviton components $h_{MN}$ defined as:
\begin{equation}
 G_{MN} = {\widetilde G}_{MN} + h_{MN}~,
\end{equation}
where
\begin{equation}
 {\widetilde G}_{MN} \equiv \exp(2A)~\eta_{MN}
\end{equation}
denotes the background de Sitter metric.
Moreover, we will use the notation $h \equiv {\widetilde G}^{MN} h_{MN}$.

{}After setting $\varphi^A = 0$, we have
\begin{eqnarray}
 && Y_{MN} = {\nu^2\over (n_S x^S)^2} ~\eta_{MN} = \mu^2 {\widetilde G}_{MN}~,\\
 && Y = Y_{MN} G^{MN} = \nu^2 H^2 \left[D - h + \dots\right] = Y_* - \mu^2 h + \dots~,
\end{eqnarray}
where the ellipses stand for higher order terms in $h_{MN}$.

{}Due to diffeomorphism invariance, the scalar equations of motion (\ref{phiY}) are related to (\ref{einsteinY}) via Bianchi identities.
We will therefore focus on (\ref{einsteinY}). Let us first rewrite it as follows:
\begin{eqnarray}
 &&R_{MN} - {1\over 2}G_{MN} \left[R - {\widetilde \Lambda}\right] = \nonumber\\
 &&\mu^2 \left[ {\widetilde G}_{MN} V^\prime(Y) - G_{MN} V^\prime(Y_*) \right] -
 {1\over 2}G_{MN}\left[V(Y) - V(Y_*)\right]~.
\end{eqnarray}
Linearizing the r.h.s. of this equation, we obtain:
\begin{eqnarray}
 R_{MN} - {1\over 2}G_{MN} \left[R - {\widetilde \Lambda}\right] = {M^2\over 2} \left[{\widetilde G}_{MN} h - \zeta h_{MN}\right] +\dots~,
\end{eqnarray}
where
\begin{eqnarray}
 &&M^2 \equiv \mu^2 V^\prime(Y_*) - 2\mu^4 V^{\prime\prime}(Y_*)~,\\
 &&\zeta M^2 \equiv 2\mu^2 V^\prime(Y_*)~.
\end{eqnarray}
This corresponds to adding a graviton mass term of the form
\begin{equation}
 -{M^2\over 4} \left[\zeta h_{MN}h^{MN} - h^2\right]
\end{equation}
to the Einstein-Hilbert action with the cosmological constant ${\widetilde \Lambda}$,
and the Fierz-Pauli combination corresponds to taking $\zeta = 1$. This occurs for a special class of potentials with
\begin{equation}\label{tune-V}
 V^\prime(Y_*) = -{2\over D} Y_* V^{\prime\prime}(Y_*)~.
\end{equation}
Thus, as we see, we can obtain the Fierz-Pauli combination of the mass term
for the graviton if we tune {\em one} combination of couplings. In fact, this tuning is nothing but the tuning of
the cosmological constant -- indeed, (\ref{tune-V}) relates the cosmological constant to higher derivative couplings.

{}Thus, consider a simple example:
\begin{equation}\label{quad.pot}
 V = \Lambda + Y + \lambda Y^2~.
\end{equation}
The first term is the cosmological constant, the second term is the kinetic term for the scalars (which can always be normalized
such that the corresponding coefficient is 1 by adjusting the coupling $\nu$), and the third term is a four-derivative term. We then have:
\begin{equation}
 Y_* = -{D\over{2(D+2)}}~\lambda^{-1}~,
\end{equation}
which relates the mass parameter $\mu$ to the higher derivative coupling $\lambda$:
\begin{equation}
 \mu^2 = Y_* / D = -{1\over{2(D+2)}}~ \lambda^{-1}~,
\end{equation}
and the graviton mass is given by:
\begin{equation}
 M^2 = -{2\over{(D+2)^2}}~ \lambda^{-1}~.
\end{equation}
Note that we must have $\lambda < 0$. Moreover, we have:
\begin{equation}\label{LL}
 {\widetilde \Lambda} = \Lambda - {{D^2 + 4D - 8}\over {4(D+2)^2}}~\lambda^{-1}~.
\end{equation}
Recall, however, that we have (\ref{Y_*}). This implies that
\begin{equation}
 {\widetilde \Lambda} = -{(D-1)(D-2)\over 2\nu^2(D+2)}~\lambda^{-1}~,
\end{equation}
and the cosmological constant $\Lambda$ needs to be tuned against the higher derivative coupling $\lambda$.

{}Finally, note that
\begin{equation}
 {{\widetilde \Lambda} \over M^2} = {(D-1)(D-2)(D+2)\over 4\nu^2}~.
\end{equation}
This ratio will become important in the next section.

\section{Enhanced Symmetry?}

{}In the previous section we saw that at the special value of the cosmological constant (given by (\ref{tune-V})) we have massive gravity in de Sitter space with the Fierz-Pauli mass term in the linearized approximation. However, as we will argue in this section, there appears to be a qualitative difference between the linearized approximation and the full nonlinear theory, at least for some values of the coupling $\nu$.

{}Before we do this, however, let us briefly comment on the counting of the propagating degrees of freedom. We started with massless gravity with $D(D - 3)/2$ propagating degrees of freedom plus $D$ scalars. However, just as in the case of massive gravity in Minkowski space via gravitational Higgs mechanism discussed in \cite{ZK1}, at the spacial value of the cosmological constant (\ref{tune-V}), due to the presence of higher derivative terms for the scalars, the kinetic term for scalar fluctuations reorganizes into that of a vector boson, and we have only $D-1$ propagating scalar degrees of freedom. These $D-1$ scalar degrees of freedom are eaten by the graviton in the process of spontaneous breaking of diffeomorphisms, the graviton acquires mass, and has $(D+1)(D-2)/2 ~(= D(D - 3)/2 + (D-1))$ propagating degrees of freedom.

{}Thus, in gravitational Higgs mechanism massive gravity arises as a result of spontaneous breaking of diffeomorphisms (as opposed to explicit breaking thereof by simply adding a mass term for the graviton fluctuations). In fact, gravitational Higgs mechanism provides a non-perturbative definition for massive gravity in the corresponding background. Once we gauge away the scalars via (\ref{diffphiY}), we obtain the following action for gravity in de Sitter background:
\begin{equation}
 S_G = M_P^{D-2}\int d^Dx \sqrt{-G}\left[ R - {\widetilde V}(G^{MN}{\widetilde G}_{MN})\right]~,
 \label{actionphiG}
\end{equation}
where we have defined ${\widetilde V}(\zeta) \equiv V(\mu^2 \zeta)$, and ${\widetilde G}_{MN}$ is the background de Sitter metric (\ref{solGY}). The equations of motion are given by
\begin{equation}
\label{einsteinG}
 R_{MN} - {1\over 2}G_{MN} R = {\widetilde V}^\prime(G^{KL}{\widetilde G}_{KL}) {\widetilde G}_{MN}
 -{1\over 2}G_{MN} {\widetilde V}(G^{KL}{\widetilde G}_{KL})~,
\end{equation}
and the Bianchi identities imply that
\begin{equation}\label{phiG.1}
 \partial_M\left[\sqrt{-G}~ {\widetilde V}^\prime(G^{KL}{\widetilde G}_{KL}) G^{MN}{\widetilde G}_{NS}\right] - {1\over 2}\sqrt{-G}~ {\widetilde V}^\prime(G^{KL}{\widetilde G}_{KL}) G^{MN}\partial_S {\widetilde G}_{MN} = 0~.
\end{equation}
Note that this condition is due to the presence of the ``mass term'' in (\ref{actionphiG}).

{}Let us now study linearized equations of motion. We expand $G_{MN} = {\widetilde G}_{MN} + h_{MN}$:
\begin{eqnarray}\label{lin.eq}
 &&\Box ~h_{MN} + \nabla_M\nabla_N h - \nabla_M\nabla^S h_{SN} - \nabla_N\nabla^S h_{SM} -
 {\widetilde G}_{MN}\left[\Box ~h - \nabla^S\nabla^R h_{SR}\right] - \nonumber\\
 &&  H^2 \left[2 h_{MN} + (D-3) {\widetilde G}_{MN} h \right] - M^2 \left[h_{MN} - {\widetilde G}_{MN} h\right] = 0~,
\end{eqnarray}
where $\nabla_M$ is the covariant derivative in the de Sitter background metric ${\widetilde G}_{MN}$, and $\Box \equiv {\widetilde G}^{MN} \nabla_M \nabla_N$. Furthermore, the condition (\ref{phiG.1}) reduces to
\begin{equation}\label{lin.con}
 \nabla^N h_{MN} - \nabla_M h = 0~.
\end{equation}
Note that (\ref{lin.con}) follows from (\ref{lin.eq}).

{}While the linearized equations of motion (\ref{lin.eq}) are not invariant under diffeomorphisms, at the special value of the ratio ${\widetilde \Lambda}^2/ M^2$ they are invariant under the following infinitesimal transformations:
\begin{equation}\label{special}
 \delta h_{MN} = (\nabla_M\nabla_N +  H^2 {\widetilde G}_{MN})\chi~.
\end{equation}
Indeed, (\ref{lin.eq}) are invariant under (\ref{special}) when
\begin{equation}\label{M-H}
 M^2 = (D-2) H^2~.
\end{equation}
The following identities are useful in deriving this result:
\begin{eqnarray}
 &&\left(\Box~\nabla_M - \nabla_M~\Box\right)\chi = (D-1) H^2~\nabla_M \chi~,\\
 &&\left(\Box~\nabla_M\nabla_N - \nabla_M\nabla_N~\Box\right)\chi = 2 H^2\left(D~\nabla_M\nabla_N - {\widetilde G}_{MN}~\Box\right)\chi~.
\end{eqnarray}
In $D=4$ the presence of the symmetry (\ref{special}) at the point (\ref{M-H}) was discussed in \cite{DN}.

{}The presence of this additional local symmetry in the linearized theory implies that at the point (\ref{M-H}) the graviton has $(D+1)(D-2)/2 - 1$ propagating degrees of freedom, one fewer than at generic points in the parameter space. Furthermore, the linearized theory is non-unitary for $M^2 < (D-2) H^2$ (as the helicity-0 graviton mode has negative norm), while for $M^2 > (D-2) H^2$ all $(D+1)(D-2)/2$ graviton modes are propagating and have positive norm \cite{DN, Higuchi, DW}. Also, in the linearized theory at $M^2 = (D-2) H^2$ the graviton $h_{MN}$ can only couple to traceless conserved energy-momentum tensor $T_{MN}$ as the coupling to the trace part of $T_{MN}$ is inconsistent with the symmetry (\ref{special}).

{}However, here we will argue that the additional local symmetry (\ref{special}) at the point (\ref{M-H}), which implies that the helicity-0 graviton mode has null norm at $M^2 = (D-2) H^2$ and further acquires negative norm for $M^2 < (D-2) H^2$, is absent in the full nonlinear theory. To see this, let us start with the de Sitter solution and transform it via (\ref{special}) with $\chi = \alpha t$, where $\alpha$ is a constant, {\em i.e.}, $\chi$ is a linear function of time $t\equiv n_S x^S$ only. For such $\chi$, the transformation (\ref{special}) reads:
\begin{equation}
 \delta h_{MN} = {2\alpha\over t} \left(n_M n_N + \eta_{MN}\right)~.
\end{equation}
This then implies that the so transformed metric is diagonal and of the form
\begin{eqnarray}\label{pert.sol.1}
 &&G_{00} = {\widetilde G}_{00}~,\\
 \label{pert.sol.2}
 &&G_{ii} = {\widetilde G}_{ii} (1 + 2 H^2\alpha t)~,
\end{eqnarray}
and is equivalent to the background de Sitter metric ${\widetilde G}_{MN}$. We will now argue that the full non-perturbative equations of motion do not possess such a symmetry or such solutions.

{}The following discussion can be straightforwardly generalized to general $V(Y)$. However, for our purposes here it will suffice to consider quadratic $V(Y)$ of the form (\ref{quad.pot}). Assuming (\ref{M-H}), from the previous section we then have:
\begin{eqnarray}
 &&4\nu^2 = (D-2)(D+2)~,\\
 &&\Lambda = -{{D(D-4)}\over{2(D+2)}}~\mu^2~,\\
 \label{V-tilde}
 &&{\widetilde V}(\zeta) = -{\mu^2\over 2(D+2)} \left[D(D-4) - 2(D+2)\zeta + \zeta^2\right]~.
\end{eqnarray}
Note that the cosmological constant $\Lambda = 0$ in $D=4$.

{}For our purposes here it will suffice to first consider (\ref{phiG.1}) as opposed to the full equations of motion (\ref{einsteinG}). We will look for solutions of the form
\begin{equation}\label{Ansatz}
 G^{MN} = {\mbox {diag}}\left({\widetilde G}^{00}, f(t)~{\widetilde G}^{ii}\right)~,
\end{equation}
where $f(t)$ is a function of time $t$ only. For such solutions, we have
\begin{equation}
 {\widetilde V}^\prime(G^{MN}{\widetilde G}_{MN}) = -{(D-1)\mu^2\over {D+2}} \left[f - {{D+1}\over {D-1}}\right]~,
\end{equation}
and (\ref{phiG.1}) reduces to the following equation for $f(t)$:
\begin{equation}
 \partial_t\left[\sqrt{-G}\left(f - {{D+1}\over {D-1}}\right)\right] + {1\over t} \sqrt{-G} \left(f - {{D+1}\over {D-1}}\right) \left[1 + (D - 1)f\right] = 0~.
\end{equation}
Using the following equation
\begin{equation}
 \partial_t \sqrt{-G} = -\sqrt{-G}\left[D + {{D - 1}\over 2} {\partial_t f \over f}\right]~,
\end{equation}
we then have:
\begin{equation}
 {1\over 2}\left[ (D + 1) - (D - 3) f\right] \partial_t f + {{D-1}\over t} f \left(f - 1\right) \left[f - {{D+1}\over {D-1}}\right] = 0~,
\end{equation}
or equivalently:
\begin{equation}\label{eq-f}
 {1\over 2}\left[{1\over f} - {2\over {f - 1}} + {1\over {f - {{D+1}\over {D-1}}}}\right] \partial_t f + {1\over t} = 0~,
\end{equation}
with the solution given by
\begin{equation}
 {(f - 1)^2 \over f\left|f - {{D+1}\over {D-1}}\right|} = (\gamma t)^2~,
\end{equation}
where $\gamma$ is an integration constant. We therefore have (here we are looking for solutions where $0<f<(D+1)/(D-1)$):
\begin{equation}\label{non-pert}
 f = \left[1 + {{D+1}\over 2(D-1)} (\gamma t)^2 + \epsilon \gamma t \sqrt{2\over{D-1}} \sqrt{1+ {(D+1)^2\over 8(D-1)} (\gamma t)^2}\right]/\left[1 + (\gamma t)^2\right]~,
\end{equation}
where $\epsilon =\pm 1$ and can be absorbed into the definition of $\gamma$: $\epsilon \gamma \rightarrow \gamma$. Note that this solution is indeed bounded: $0<f<(D+1)/(D-1)$.

{}If we linearize (\ref{eq-f}) via $f = 1 + \psi$, then we obtain the linearized solution (\ref{pert.sol.2}). The solution (\ref{non-pert}) is therefore the non-perturbative counterpart of the perturbative solution (\ref{pert.sol.2}) with the identification $\gamma = -\sqrt{2(D-1)} H^2\alpha$. However, we will now show that the non-perturbative solution (\ref{non-pert}) satisfies the full equations of motion (\ref{einsteinG}) only for $\gamma = 0$.

{}To see this, it will suffice to consider the trace part of the equations of motion (\ref{einsteinG}):
\begin{equation}
 R = {1\over{D-2}}\left[D{\widetilde V}(\zeta) - 2\zeta{\widetilde V}^\prime(\zeta)\right]~,
\end{equation}
where $\zeta\equiv G^{MN}{\widetilde G}_{MN} = 1 + (D-1)f = D + (D-1)\psi$. Using (\ref{V-tilde}) we then have:
\begin{eqnarray}
 R &=&{ H^2 \over 8}\left[(4-D) \zeta^2 + 2(D-2)(D+2)\zeta + D^2(4-D)\right] = \nonumber\\
 &&D(D-1) H^2\left[1 + {{D-1}\over D}\psi + {(D-1)(4-D)\over 8 D}\psi^2\right]~.
\end{eqnarray}
In $D=4$ we have a simplification; however, we will continue to work in general $D$.

{}From the definition of the Ricci scalar, we have:
\begin{equation}
 R = D(D - 1) H^2\left[1 + \partial_* \ln(f) - {1\over D} \partial^2_* \ln(f) + {1\over 4} \left(\partial_*\ln(f)\right)^2\right]~,
\end{equation}
where $\partial_* \equiv t\partial_t$. We therefore have the following equation of motion for $\psi = f - 1$:
\begin{equation}\label{eq-psi}
 \partial_* \ln(f) - {1\over D} \partial^2_* \ln(f) + {1\over 4} \left(\partial_*\ln(f)\right)^2 =
 {{D-1}\over D}\psi + {(D-1)(4-D)\over 8 D}\psi^2~.
\end{equation}
Note that (\ref{eq-psi}) is exact.

{}Note that the linearized version of (\ref{eq-psi})
\begin{equation}
 \partial_* \psi - {1\over D} \partial^2_* \psi = {{D-1}\over D}\psi
\end{equation}
admits solutions $\psi = \xi t$, where $\xi$ is an integration constant, which match such linear solutions of the linearized version of (\ref{non-pert}). However, this does not hold beyond the linearized level. Indeed, it is not difficult to show that (\ref{non-pert}) does not satisfy (\ref{eq-psi}) except for $\gamma = 0$. A simple way to see this is to solve (\ref{eq-psi}) to the second order in $t$ and compare this solution to (\ref{non-pert}) expanded to the same order:
\begin{eqnarray}
 &&{\mbox {Eq.} (\ref{non-pert}):}~~~\psi_1 = {\widetilde \gamma} t + {{3-D}\over 4}({\widetilde \gamma}t)^2 + {\cal O}({\widetilde \gamma} t)^3~,\\
 &&{\mbox {Eq.} (\ref{eq-psi}):}~~~\psi_2 = \eta t - {{D^2 - 11 D + 20}\over {8 (D - 3)}}(\eta t)^2 + {\cal O}(\eta t)^3~,
\end{eqnarray}
where we have defined ${\widetilde \gamma} \equiv \sqrt{(D-1)/2} ~\gamma$, and $\eta$ is an integration constant. As we see, the two solutions do not match. Once we identify $\eta = {\widetilde \gamma}$, the difference between the two solutions is given by:
\begin{equation}
 \psi_2 - \psi_1 = {(D+1)(D-2)\over 8(D-3)}({\widetilde \gamma}t)^2 + {\cal O}({\widetilde \gamma} t)^3~.
\end{equation}
This implies that the full non-perturbative equations of motion (\ref{einsteinG}) do not admit solutions of the form (\ref{Ansatz}) (except for $f\equiv 1$).

{}An intuitive way of seeing why the linearized theory possesses the additional symmetry not present in the full nonlinear theory is as follows. In the linearized theory, (\ref{phiG.1}) reduces to (\ref{lin.con}), which in turn implies that the part of the Einstein tensor containing covariant derivatives is traceless. The remaining part then is traceless at the special point (\ref{M-H}). The tracelessness of the Einstein tensor is what leads to the appearance of the enhanced local symmetry in the linearized theory. Indeed, as we mentioned above, a conserved energy-momentum tensor for matter sources is compatible with this enhanced symmetry only if it is traceless.

{}On the other hand, in the full nonlinear theory (\ref{phiG.1}), unlike (\ref{lin.con}), depends on the structure of the potential ${\widetilde V}$, and therefore it cannot possibly make the derivative part of the Einstein tensor traceless as the latter intrinsically knows nothing about the structure of ${\widetilde V}$. Therefore, non-perturbatively, one does not expect to have an enhanced local symmetry that would remove a propagating degree of freedom. Indeed, for generic potentials ${\widetilde V}$, the action (\ref{actionphiG}) does not possess any local symmetries.

{}Let us quantify the previous two paragraphs. First, note that the scalars $\phi^A$ (or, their fluctuations around the background (\ref{solphiY}) and (\ref{solGY})) constitute the matter fields in the action (\ref{actionphiY}). The energy-momentum tensor reads:
\begin{equation}
 T_{MN} = -2 M_P^{D-2} \sqrt{G} V^\prime(Y) Y_{MN}~,
\end{equation}
which is not traceless in the Higgs phase. Also, a nonlinear completion of (\ref{special}), {\em before} gauging away the scalars, is given by
\begin{equation}
 \delta G_{MN} = \nabla_M \nabla_N \chi + {1\over \nu^2} Y_{MN} \chi~,
\end{equation}
where $\nabla_M$ is the covariant derivative in the metric $G_{MN}$. In the Higgs phase, once we gauge away the scalars, we have
\begin{equation}\label{special-nonpert}
 \delta G_{MN} = \nabla_M \nabla_N \chi +  H^2 {\widetilde G}_{MN} \chi~,
\end{equation}
where $\nabla_M$ is the covariant derivative in the metric $G_{MN}$. However, (\ref{special-nonpert}) is not a symmetry of the full nonlinear action (\ref{actionphiG}).

{}Here the following remark is in order. In (\ref{Ansatz}) we assume that $G^{00} = {\widetilde G}^{00}$. Here we can ask if there exist more general solutions with $G^{00} = g(t) {\widetilde G}^{00}$ (and $G^{ii} = f(t) {\widetilde G}^{ii}$) matching the perturbative solutions (\ref{pert.sol.1}) and (\ref{pert.sol.2})\footnote{Based on symmetry considerations, namely, the $SO(D-1)$ invariance in the spatial directions, the off-diagonal terms in $G^{MN}$ are not relevant in this discussion.}. For this to be the case, we must have
\begin{equation}
 g = 1 + {\cal O}(\eta t)^2~.
\end{equation}
However, if the symmetry (\ref{special}) is indeed present, then we can always transform the metric such that $g = 1 + {\cal O}(\eta t)^3$. Indeed, for general $\chi(t)$ that depends on time $t$ only, (\ref{special}) reads:
\begin{equation}
 \delta h_{MN} = n_M n_N \left[\partial^2_t \chi + {2\over t}\partial_t\chi\right] +  H^2 \left[t\partial_t \chi + \chi\right]{\widetilde G}_{MN}~.
\end{equation}
This implies that any ${\cal O}(\eta t)^2$ term in $g$ can be transformed away by including an appropriate ${\cal O}(\eta t)^2$ term in $\chi$. More concretely, if $g = 1 + a (\eta t)^2 + {\cal O}(\eta t)^3$, then the transformation (\ref{special}) with $\chi = a\eta^2/3 H^2$ will result in $g = 1 + {\cal O}(\eta t)^3$. Therefore, if the enhanced local symmetry (\ref{special}) were present, our assumption of $G^{00} = {\widetilde G}^{00}$ would hold.

\section{Is there a Ghost?}

{}In the previous section we argued that in the full nonlinear theory there is no enhanced local symmetry at $M^2 = (D-2) H^2$. Absent such local symmetry, one does not expect appearance of a null norm state at $M^2 = (D-2) H^2$, which would make it difficult to imagine that a negative norm state would appear for $M^2 < (D-2) H^2$. This suggests that no ghost might be present for lower graviton mass values and the theory might be unitary for all values of the graviton mass $M$ and the Hubble parameter $ H$ with no vDVZ discontinuity.\footnote{The related issue of causality will be relegated to future work. The possibility of violating causality by means of superluminal propagation of signals, which affects other models of modified gravity, cannot be checked easily in the model at hand. Namely, the phase velocity at low wavelengths of a linearized perturbation, the speed of propagation of a signal, has no meaning in a model where, as argued above, the linearized analysis is not applicable, and the study of the corresponding phenomenon at nonlinear level is beyond the scope of this article.} The purpose of this section is to argue that this is indeed the case. We will do this by studying the full nonlinear action for the relevant modes, which we identify next. In particular, we will argue that no negative norm state is present for any value of the ratio $M/ H$.

{}To identify the relevant modes in the full nonlinear theory, let us note that in the linearized theory the potentially ``troublesome'' mode is the longitudinal helicity-0 mode $\rho$. However, we must also include the conformal mode $\omega$ as there is kinetic mixing between $\rho$ and $\omega$. In fact, $\rho$ and $\omega$ are not independent but are related via Bianchi identities. Therefore, in the linearized language one must look at the modes of the form
\begin{equation}\label{param.lin}
 h_{MN} = {\widetilde G}_{MN}~\omega + \nabla_M\nabla_N \rho~.
\end{equation}
Furthermore, based on symmetry considerations, namely, the $SO(D-1)$ invariance in the spatial directions, we can focus on field configurations independent of spatial coordinates. Indeed, for our purposes here we can compactify the spatial coordinates on a torus $T^{D-1}$ and disregard the Kaluza-Klein modes. This way we reduce the $D$-dimensional theory to a classical mechanical system, which suffices for our purposes here. Indeed, with proper care (see below), if there is a negative norm state in the uncompactified theory, it will be visible in its compactified version, and vice-versa.

{}Let us therefore consider field configurations of the form:
\begin{equation}\label{param.full}
 G^{MN} = {\rm diag}(g(t)~{\widetilde G}^{00}, f(t)~{\widetilde G}^{ii})~,
\end{equation}
where $g(t)$ and $f(t)$ are functions of time $t$ only. The action (\ref{actionphiG}) then reduces as follows:
\begin{eqnarray}\label{compact}
 S_G = -\kappa \int {dt\over t^D} ~g^{-{1\over 2}} f^{-{{D-1}\over 2}}\left\{{\widetilde \Lambda}g U^2 + {\widetilde V}(g + \Omega)\right\}~,
\end{eqnarray}
where
\begin{eqnarray}
 &&\kappa\equiv {M_P^{D-2} W_{D-1}\over  H^D}~,\\
 &&U\equiv 1 + {1\over 2}\partial_*\ln(f)~,\\
 &&\Omega\equiv (D-1)f~,
\end{eqnarray}
and $W_{D-1}$ is the volume in the spatial dimensions ({\em i.e.}, the volume of $T^{D-1}$). Note that $g$ is a Lagrange multiplier. The goal is to integrate out $g$ and obtain the corresponding action for $f$. It is then this action that we should test for the presence of a negative norm state.

{}The equation of motion for $g$ reads:
\begin{equation}
 {\widetilde V}(g + \Omega) - 2g{\widetilde V}^\prime(g + \Omega) = {\widetilde \Lambda}g U^2~.\label{geq}
\end{equation}
The following discussion can be straightforwardly generalized to general ${\widetilde V}$. However, for our purposes here it will suffice to consider quadratic ${\widetilde V}$ corresponding to (\ref{quad.pot}). We then have:
\begin{equation}\label{eqn.g}
 3\lambda\mu^2g^2 + \left[1 + 2\lambda\mu^2\Omega + {{\widetilde\Lambda}\over{\mu^2}}~U^2\right] g -
 \left\{{\Lambda\over \mu^2} + \Omega\left[1 + \lambda\mu^2\Omega\right]\right\} = 0~.
\end{equation}
We can therefore express $g$ in terms of $f$ and $\partial_* \ln(f)$:
\begin{eqnarray}
 6\lambda\mu^2 g = &&-\left[1 + 2\lambda\mu^2\Omega + {{\widetilde \Lambda}\over\mu^2}~U^2\right] +\nonumber\\
 &&\sqrt{\left[1 + 2\lambda\mu^2\Omega + {{\widetilde \Lambda}\over\mu^2}~U^2\right]^2 + 12\lambda\mu^2\left\{{\Lambda\over \mu^2} + \Omega\left[1 + \lambda\mu^2\Omega\right]\right\}}~,\label{sol.g}
\end{eqnarray}
where the branch is fixed by the requirement that $g \equiv 1$ when $f \equiv 1$. Substituting the so expressed $g$ into (\ref{compact}), we obtain an action which is a nonlinear functional of $f$ and $\partial_* \ln(f)$.

{}For our purposes here it is more convenient to work with the logarithmic time coordinate $\tau$ and the canonical variable $q$, where
\begin{eqnarray}
 &&\tau\equiv \ln( H t)~,\\
 &&q\equiv \ln(f) + 2\tau~,\\
 &&\Omega = (D-1)e^{q - 2\tau}~,\\
 &&U = {1\over 2} \partial_\tau q~,
\end{eqnarray}
and the action reads:
\begin{eqnarray}\label{compact1}
 S_G = \int d\tau L = -\kappa H^{D-1} \int {d\tau} ~g^{-{1\over 2}} e^{-{{D-1}\over 2}q}\left\{{\widetilde \Lambda} g U^2 + {\widetilde V}(g + \Omega)\right\}~,
\end{eqnarray}
where $L$ is the Lagrangian. This action corresponds to a classical mechanical system with a lagrange multiplier $g$ and a time-dependent potential. Upon integrating out the Lagrange multiplier, the time dependence also propagates into the ``kinetic'' (or, more precisely, momentum-dependent) terms.

{}Next, the conjugate momentum is given by
\begin{eqnarray}
 &&p = {{\partial L} \over {\partial(\partial_\tau q)}} = -\kappa H^{D-1} e^{-{{D-1}\over 2}q} \times \nonumber\\
 && \times \left\{{1\over 2} g^{-{1\over 2}} {\hat g} {\widetilde \Lambda} U^2 - {1\over 2} g^{-{3\over 2}} {\hat g} {\widetilde V}(g + \Omega) + g^{-{1\over 2}} {\hat g} {\widetilde V}^\prime(g + \Omega) + g^{{1\over 2}} {\widetilde \Lambda} U\right\}~,\label{momentum}
\end{eqnarray}
where
\begin{equation}
 {\hat g} \equiv {{\partial g} \over {\partial(\partial_\tau q)}}~.
\end{equation}
Using (\ref{geq}), (\ref{momentum}) simplifies to
\begin{equation}
 p = -\kappa H^{D-1} e^{-{{D-1}\over 2}q} g^{{1\over 2}} {\widetilde \Lambda} U~,
\end{equation}
and the Hamiltonian is given by
\begin{equation}
 {\cal H} = p~\partial_\tau q - L = -\kappa H^{D-1} g^{-{1\over 2}} e^{-{{D-1}\over 2}q} \left[{\widetilde \Lambda} g U^2 - {\widetilde V}(g + \Omega)\right]~.
\end{equation}
We can now see if this Hamiltonian is bounded from below.

{}First, using (\ref{geq}), we have:
\begin{equation}
 {\cal H} = 2 \kappa H^{D-1} g^{{1\over 2}} e^{-{{D-1}\over 2}q} {\widetilde V}^\prime(g + \Omega) =
 2 \mu^2 \kappa H^{D-1} g^{{1\over 2}} e^{-{{D-1}\over 2}q}\left[1 + 2\lambda\mu^2(g + \Omega)\right]~.
\end{equation}
Using (\ref{sol.g}), we can rewrite this Hamiltonian as follows:
\begin{equation}
 {\cal H} = {2\over 3} \mu^2 \kappa H^{D-1} g^{{1\over 2}} e^{-{{D-1}\over 2}q}\left[X - Z\right]~,
\end{equation}
where
\begin{eqnarray}
 &&X \equiv \sqrt{\left[1 + 2\lambda\mu^2\Omega + {{\widetilde \Lambda}\over\mu^2}~U^2\right]^2 + 12\lambda\mu^2\left\{{\Lambda\over \mu^2} + \Omega\left[1 + \lambda\mu^2\Omega\right]\right\}}~,\\
 &&Z\equiv {{\widetilde \Lambda}\over\mu^2}~U^2 - 4\lambda\mu^2\Omega - 2~.
\end{eqnarray}
The presence of a ghost would imply that the Hamiltonian is unbounded from below for large values of $U^2$ (recall that $U^2$ contains the ``kinetic'' term). However, it is not difficult to show that this Hamiltonian suffers from no such pathology. Indeed, we can rewrite it as follows:
\begin{eqnarray}
 {\cal H} &=& {2\over 3} \mu^2 \kappa H^{D-1} g^{{1\over 2}} e^{-{{D-1}\over 2}q}~{{X^2 - Z^2}\over {X + Z}} = \nonumber\\
 &&2 \mu^2 \kappa H^{D-1} g^{{1\over 2}} e^{-{{D-1}\over 2}q}~ {{4\lambda\Lambda - 1 + 2 ({\widetilde \Lambda} / \mu^2)~U^2
 \left[1 + 2\lambda\mu^2 \Omega\right]} \over {X + Z}}~,
\end{eqnarray}
which in the large $U^2$ limit reads:
\begin{equation}
 {\cal H} = 2 \mu^2 \kappa H^{D-1} g^{{1\over 2}} e^{-{{D-1}\over 2}q}\left[1 + 2\lambda\mu^2 \Omega\right] +{\cal O}(1/U^2)~.
\end{equation}
Furthermore, from (\ref{sol.g}) we have
\begin{equation}
 6\lambda\mu^2 g = {{X^2 - Q^2} \over{X + Q}} = 12 ~{\lambda\Lambda + \lambda\mu^2\Omega\left[1 + \lambda\mu^2\Omega\right] \over{X + Q}} = {\cal O}(1/U^2)~,
\end{equation}
where
\begin{equation}
 Q\equiv {{\widetilde \Lambda}\over\mu^2}~U^2 + 2\lambda\mu^2\Omega + 1~.
\end{equation}
So, in the large $U^2$ limit the Hamiltonian actually vanishes.

{}Note that the above argument implicitly assumes that $\Omega$ is bounded from above. This is indeed the case as $g$ must be at least non-negative, which implies that
\begin{equation}
 \lambda\Lambda + \lambda\mu^2\Omega\left[1 + \lambda\mu^2\Omega\right] \leq 0~,
\end{equation}
and $\Omega$ is bounded as follows (note that we must have $\Omega \geq 0$):
\begin{equation}
 \mbox{max}\left(0~,~ -{{1 - \sqrt{1 - 4\lambda\Lambda}}\over 2\lambda\mu^2}\right) \leq \Omega \leq -{{1 + \sqrt{1 - 4\lambda\Lambda}}\over 2\lambda\mu^2}~,
\end{equation}
and we must further have
\begin{equation}
 \Lambda \geq {1\over 4\lambda}~,
\end{equation}
which together with (\ref{LL}) implies that we must have
\begin{equation}
 {\widetilde \Lambda} \geq {3 \over (D+2)^2}\lambda^{-1}~,
\end{equation}
which is always satisfied as $\lambda < 0$.

{}Thus, as we see, there appears to be no ghost in the full nonlinear theory. To understand why a ghost is present in the linearized theory, let us linearize our Hamiltonian. To do this, we will assume that $f = 1 + \psi$, where $|\psi| \ll 1$, {\em i.e.}, we consider small fluctuations around the de Sitter background. Furthermore, we also assume that $\left|\partial_\tau \psi\right|\ll 1$, so we can linearize the Hamiltonian to the second order in $\partial_\tau \psi$ as well as $\psi$. In fact, here we are interested in the terms containing $(\partial_\tau \psi)^2$. A straightforward computation gives the following linearized Hamiltonian (the ellipses stand for the terms not containing $(\partial_\tau \psi)^2$):
\begin{equation}
 {\cal H}_* = -{3\over 2} \mu^2 \kappa H^{D-1} e^{-(D-1)\tau}\left(\partial_\tau \psi\right)^2 + \dots~,
\end{equation}
{\em i.e.}, this linearized Hamiltonian contains a ghost for all values of $M$ and $ H$, which is absent in the full nonlinear theory. In this regard, we discuss a simple illustrative example in Appendix \ref{A}.

{}Here one might find it puzzling that, all the potential pitfalls of linearization notwithstanding, in the linearized theory the ghost appears for all values of $M$ and $ H$, while according to \cite{DN, Higuchi, DW} a negative norm state is expected to appear only for $M^2 < (D-2) H^2$.  The difference here is due to the parametrization of the conformal and helicity-0 longitudinal modes. We have been working with (\ref{param.full}), while the aforesaid result of \cite{DN, Higuchi, DW} applies to (\ref{param.lin}). The difference between the two is that (\ref{param.full}) has no derivatives. In this regard, one might wonder if the ghost is ``masked'' by (\ref{param.full})\footnote{Consider a ghost in $D$ dimensions: $L = -c ~\partial^M\phi\partial_M\phi$, $c < 0$. Compactify on $T^{D-1}$: $L = c_1 (\partial_t\phi)^2$ ($c_1 \equiv c W_{D-1}$, $W_{D-1}$ being the volume of $T^{D-1}$). The Hamiltonian is not bounded from below: ${\cal H} = p^2/4c_1$, where the conjugate momentum $p = 2c_1\partial_t\phi$. However, let $q \equiv \partial_t\phi$. If $q$ is treated naively as the canonical variable, then $L = c_1 q^2$, ${\cal H} = -c_1 q^2$, which is bounded from below. That is, the ghost appears to have been ``masked'' by redefining the variables. The flaw in the argument is that the transformation of variables is not a canonical one, therefore, the new Hamiltonian describes a different dynamical system.}. Furthermore, one might wonder if the dimensionally reduced action (\ref{compact1}) provides an adequate description.

{}In this regard, we have explicitly checked that if we expand the action (\ref{compact1}) to the quadratic order in the parametrization corresponding to (\ref{param.lin}), we obtain that there is a null norm state at $M^2 = (D-2) H^2$ and a ghost at $M^2 < (D-2) H^2$, so the dimensionally reduced action (\ref{compact1}) correctly reproduces the linearized results \cite{DW2} in the parametrization corresponding to (\ref{param.lin}). In fact, what transpires is the following. At the quadratic order, the second derivatives introduced by the parametrization corresponding to (\ref{param.lin}) can be integrated by parts to arrive at an action containing only first derivatives of $\omega$ and $\rho$. This action then possesses the aforesaid properties w.r.t. the appearance of a null norm state and a ghost. However, we have explicitly checked that already at the cubic level the second derivatives introduced by the parametrization corresponding to (\ref{param.lin}) cannot be integrated by parts, so the resulting action invariably includes terms with second derivatives of $\rho$. This is clearly problematic already at the cubic level and suggests that the parametrization corresponding to (\ref{param.lin}) cannot be used beyond the linearized approximation.

{}Indeed, according to (\ref{param.lin}) the ghost would appear for $M^2 < (D-2) H^2$, while at $M^2 = (D-2) H^2$ we would have a null norm state, and for $M^2 > (D-2) H^2$ the theory is unitary. The appearance of a null norm state would signal the presence of an enhanced local symmetry at $M^2 = (D-2) H^2$. However, in Section 4 we saw that there is no such enhanced symmetry in the full nonlinear theory. Therefore, in the full nonlinear theory we either have a ghost for all values of $M$ and $ H$, or the theory is unitary\footnote{In this regard, note that, on general grounds, for $M^2 \gg  H^2$ one expects no ghost to be present.} for all values of $M$ and $ H$. This suggests that the ``special'' point $M^2 = (D-2) H^2$ arises in the linearized theory both due to linearization and the parametrization (\ref{param.lin}). Indeed, a nonlinear completion of (\ref{param.lin}) is given by:
\begin{equation}
 G^{MN} = {\widetilde G}^{MN} f + \nabla^M\nabla^N u~.
\end{equation}
where the covariant derivative is defined w.r.t. the metric ${\widetilde G}_{MN}$ (this choice does not affect our discussion here). Note, however, that such a parametrization of the metric is rather problematic in the context of the full nonlinear theory as it introduces higher derivative terms in $u$, which should therefore not be used as the canonical variable in the full nonlinear theory. This suggests that our parametrization (\ref{param.full}) is indeed adequate\footnote{In fact, without giving details, let us simply mention that, one arrives at the same conclusion by adding a Lagrange multiplier $\eta$ leading to a constraint $g = f -  H^2 \partial_\tau^2 u$, which is a nonlinear completion of (\ref{param.lin}), and which follows from the following additional term in the action: $-\kappa H^{D-1}\int \tau \left\{ \eta \left[g - f\right] - H^2\partial_\tau\eta\partial_\tau u \right\}$.}. As we saw, in this parametrization there is a ghost in the linearized theory for all\footnote{In this regard, there is no ghost ``masking'' as a ghost does arise in this parametrization upon linearization.} values of $M$ and $H$. However, the full nonlinear theory appears to be unitary for all values of $M$ and $ H$.

{}Before concluding, let us comment on another related issue. In the linearized theory, a ``fifth constraint'' (in the $D=4$ language), which is complementary to the Bianchi identities, removes one degree of freedom (see, for example, \cite{Porrati:2000cp} for a derivation). At the special point $M^2 = (D-2) H^2$, this constraint trivializes and leads to an enhanced symmetry which removes yet another degree of freedom - as mentioned above, the norm of one of the modes becomes null.  At the nonlinear level, however, these features disappear since extra momentum and position dependent terms also contribute to the would-be constraint. In this regard, one might worry whether a propagating (ghostlike) ``sixth mode'' (in the $D=4$ language) might also be present. However, our non-perturbative Hamiltonian analysis appears to indicate that no ghostlike states or instabilities are present in the full nonlinear theory.

\section{The Upshot}

{}As we argued in Section 4, in the full nonlinear theory there is no enhanced local symmetry at the special point (\ref{M-H}). In particular, we argued that the full nonlinear theory does not admit solutions obtained by transforming the de Sitter metric via such enhanced local symmetry transformations.

{}Absent such enhanced local symmetry at the special point (\ref{M-H}) in the full nonlinear theory, we do not expect to have a null norm state and a reduction in the number of propagating degrees of freedom. Furthermore, there is also no reason to believe that for $M^2 < (D-2) H^2$ there is a negative norm state in the full nonlinear theory. If so, we can expect that, in the context of gravitational Higgs mechanism, where diffeomorphisms are broken spontaneously, there should be no van Dam-Veltman-Zakharov \cite{vDV,Zak} discontinuity, absence of which in the context of massive gravity via gravitational Higgs mechanism in Minkowski space was argued in \cite{ZK2}. In fact, from our analysis of the full nonlinear Hamiltonian for the relevant conformal and helicity-0 longitudinal modes it indeed appears that there is no ghost in the full nonlinear theory. In this regard, it appears that the vacuum corresponding to the linearized theory is just another vacuum that is unstable, and that the true vacuum corresponding to the full nonlinear theory is a different and, apparently, stable vacuum. Here one {\em analogy} that comes to mind is ghost condensation: An approximated version of a theory appears to have a ghost, but in the true vacuum the ghost condenses and the theory lacks such pathologies.

{}The upshot is that we have presented evidence that gravitational Higgs mechanism may provide a non-perturbative definition of massive gravity (both in Minkowski and de Sitter backgrounds, as well as in general curved backgrounds \cite{Curved}), and it appears that massive gravity obtained via gravitational Higgs mechanism, on general grounds, may be consistent since diffeomorphisms in gravitational Higgs mechanism are broken spontaneously (as opposed to explicit breaking by simply adding the Fierz-Pauli term for the graviton). This may open a new arena for studying infrared modified (that is, massive) gravity in the context of cosmology with non-vanishing cosmological constant.

\section*{Acknowledgements}
The work of AI is supported by the Humboldt-Foundation. AI would also like to thank Gia Dvali for valuable comments on the manuscript.

\appendix

\section{Some Pitfalls of Linearization}\label{A}

{}In this appendix we illustrate some pitfalls of linearization. Namely, apparently a nonlinear theory can have a bounded-from-below Hamiltonian, while a linearized version thereof can have a ghost. Consider the following simple toy Lagrangian for a single scalar field $\phi$:
\begin{equation}\label{toy}
 L = c \sqrt{Y^2 + 2 a Y + b^2}~,
\end{equation}
where
\begin{equation}
 Y \equiv \partial_M\phi \partial^M\phi~,
\end{equation}
$b > 0$, and $a$ can be either positive, negative or zero. To make sure that the square root in (\ref{toy}) is well defined, let us assume that
\begin{equation}\label{good.sqrt}
 |a| < b~.
\end{equation}
Next, suppose we naively linearize (to the quadratic order in $\phi$ or, equivalently, the first order in $Y$). The so linearized Lagrangian reads:
\begin{equation}
 L_* = c b + (c a/b) Y + {\cal O}(Y^2)~,
\end{equation}
which has a ghost for
\begin{equation}
 c a > 0~.\label{ghost}
\end{equation}
However, the full nonlinear theory does not possess a ghost. To see this, let us first compactify the spatial directions on a $(D-1)$-dimensional torus $T^{D-1}$ and reduce the theory to a classical mechanical system. The Lagrangian now reads:
\begin{equation}
 L = c_1 \sqrt{Z^2 - 2 a Z + b^2}~,
\end{equation}
where $c_1\equiv c W_{D-1}$, $W_{D-1}$ is the volume of $T^{D-1}$, and
\begin{equation}
 Z \equiv (\partial_t\phi)^2
\end{equation}
The conjugate momentum is given by
\begin{equation}
 p = {2c_1(Z - a) (\partial_t\phi)\over \sqrt{Z^2 - 2 a Z + b^2}}~,
\end{equation}
and the Hamiltonian reads:
\begin{equation}
 {\cal H} = {c_1 (Z^2 - b^2) \over \sqrt{Z^2 - 2 a Z + b^2}}~.
\end{equation}
Because of (\ref{good.sqrt}), the square root is well defined. Furthermore, for $c > 0$ ({\em i.e.}, $c_1 > 0)$ the Hamiltonian is bounded from below regardless of the sign of $a$. Therefore, there is no ghost in the full nonlinear theory, and its appearance in the linearized version thereof is due to linearization.


\end{document}